# Nano-photocurrent mapping of local electronic structure in twisted bilayer graphene

Sai S. Sunku[1,2]*, Alexander S. McLeod[1,]*, Tobias Stauber[3], Hyobin Yoo[4], Dorri Halbertal[1], Guangxin Ni[1], Aaron Sternbach[1], Bor-Yuan Jiang[5], Takashi Taniguchi[6], Kenji Watanabe[6], Philip Kim[4], Michael M. Fogler[5], Dimitri N. Basov[1,]†

1. Department of Physics, Columbia University, New York, NY

2. Department of Applied Physics and Applied Mathematics, Columbia University, New York, NY

3. Departamento de Teoría y Simulación de Materiales, Instituto de Ciencia de Materiales de Madrid, CSIC, Madrid, Spain

4. Department of Physics, Harvard University, Cambridge, MA

5. Department of Physics, UC San Diego, La Jolla, CA

6. National Institute for Materials Science, Tsukuba, Japan

\* - These authors contributed equally

† - Corresponding author db3056@columbia.edu

We report a combined nano-photocurrent and infrared nanoscopy study of twisted bilayer graphene (TBG) enabling access to the local electronic phenomena at length scales as short as 20 nm. We show that the photocurrent changes sign at carrier densities tracking the local superlattice density of states of TBG. We use this property to identify domains of varying local twist angle by local photo-thermoelectric effect. Consistent with the photocurrent study, infrared nano-imaging experiments reveal optical conductivity features dominated by twist-angle dependent interband transitions. Our results provide a fast and robust method for mapping the electronic structure of TBG and suggest that similar methods can be broadly applied to probe electronic inhomogeneities of moiré superlattices in other van der Waals heterostructures.

The relative twist angle $\theta$ between proximal atomic layers is emerging as an extremely capable control parameter in van der Waals (vdW) heterostructures including twisted bilayer graphene (TBG). The twist leads to a spatial variation of the atomic stacking of proximal layers with the period given by $\lambda_M = 0.246$ nm$/(2 \sin(\theta/2))$ (*1*). The resultant structure is referred to as a moiré superlattice. The electronic structure of such a superlattice consists of a large number of minibands (*2*, *3*), exhibiting strong $\theta$-dependent Van Hove singularities in the density of states (*1*, *4*, *5*). When $\theta$ is close to $1.1°$ ($\lambda_M \approx 13$nm), the lowest energy minibands in TBG specimens become nearly flat. At such "magic angle", TBG is found to host unconventional correlated electronic phases (*6*–*9*).

The electronic structure of TBG is not spatially uniform. Within the Moiré unit cell, changes in the atomic stacking lead to differences in the local density of states that have been observed by scanning tunneling microscopy (STM) experiments (*10*–*13*). At small twist angles, where atomic relaxation leads to a periodic array of topologically protected states (*14*, *15*), scanning nano-infrared (*16*) and STM (*17*) experiments have directly visualized such states. At larger length scales, variations in $\theta$ itself have been observed with multiple techniques. Low temperature transport experiments carried out as a function of the carrier density reveal a drop in conductivity when the four lowest energy minibands (which are nearly spin-valley degenerate) are completely filled (*5*, *18*) and the chemical potential reaches the superlattice band edge (SBE). The carrier density at the SBE, $n_s = 8/(\sqrt{3}\, \lambda_M^2)$, is governed by the superlattice period, and so, by $\theta$. In such transport experiments done on different parts of the same TBG device, the drop in conductivity appeared at different carrier densities, indicating a change in $\theta$ across the device (*19*). STM (*10*, *12*), transmission electron microscopy (TEM) (*15*), scanning superconducting quantum interference device (SQUID) (*20*) and scanning single electron transistor (SET) (*21*) experiments have confirmed that the electronic structure variations persist down to sub-micron scales. Moreover, experiments on magic-angle TBG suggest that reducing such fluctuations can reveal the intrinsic transport properties of TBG (*22*). Taken together, these observations suggest that understanding the variations in electronic structure of TBG on a nanometer length scale is crucial.

Here we report that a nascent optoelectronic probe, scanning photocurrent nanoscopy (*23*–*25*), can map the DC conductivity of TBG as a function of carrier density, and thereby its twist angle, with a resolution better than 20 nm. In our experiments, the DC photo-generated current across the device is measured as a function of the position of a sharp metallized tip (Fig. 1A) (*23*). Room-temperature photocurrent imaging has been previously used to study excitations such as plasmon polaritons (*24*, *25*) and phonon polaritons (*26*) in other van der Waals materials. A crucial new element of our approach was performing the experiments at cryogenic temperatures, which allowed us to visualize the insulating states in TBG (*5*, *18*). We further augmented our photocurrent results with room temperature nano-infrared imaging experiments. In tandem, our infrared and photocurrent data lead to a consistent interpretation of all the observables in terms of TBG twist-angle domains.

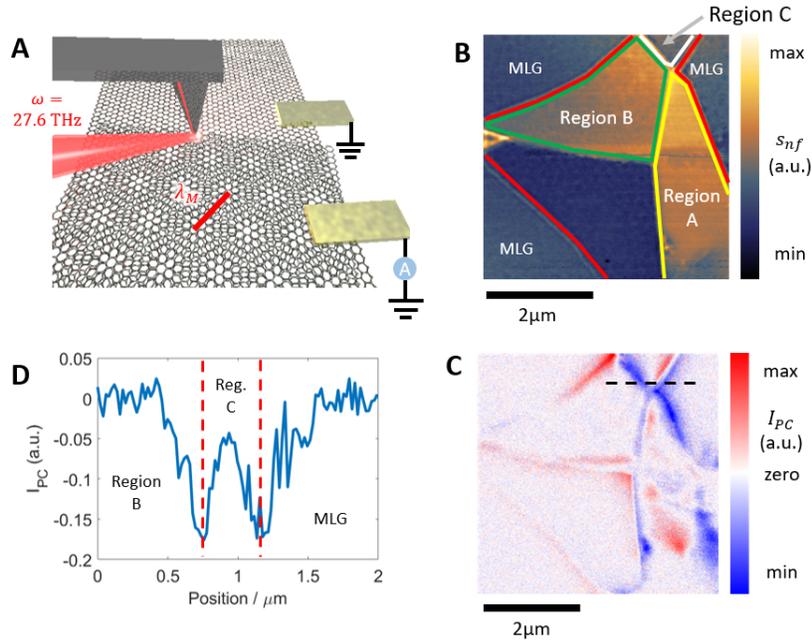

**Figure 1 | Nano-photocurrent imaging of multi-domain twisted bilayer graphene.** (A) Schematic of the experimental setup. The periodicity of the moiré pattern in TBG is denoted by $\lambda_M$. (B) Nano-infrared amplitude $s_{nf}$ image showing contrast between TBG domains. The red lines enclose the TBG region. Scale bar 2µm. (C) Nano-photocurrent image of the same region as (B) at $T = 300$K. Scale bar 2µm. (D) Line profile of nano-photocurrent across the dashed line in (C) at $T = 300$K.

Our nano-photocurrent experiments rely on the photothermoelectric effect, the dominant mechanism for photocurrent generation in graphene layers (*27, 28*). At the boundary between MLG and TBG, which we studied in this work, the magnitude of photocurrent is proportional to the difference in Seebeck coefficients of MLG and TBG. By exploiting this relationship, we determined the twist angle of the TBG region immediately adjacent to the boundary. We then performed carrier density dependent nano-infrared measurements which validated our nano-photocurrent results and provided further insight into the interband transitions that dominate the optical properties of TBG.

Our TBG structures were fabricated using the standard tear-and-stack method. The full devices included a thin top layer of hBN, the TBG, the SiO$_2$ substrate, Si gate, and electrical contacts (see Methods). The lack of an underlying hBN layer led to a high carrier density of $-1.8 \cdot 10^{13}$cm$^{-2}$ (determined through Raman spectroscopy and plasmon wavelength measurements on monolayer graphene, Section S2 of (*29*)) even when no bias was applied to the back gate, $V_G = $ 0V. We first performed room-temperature infrared nanoscopy (Fig. 1A), where we focused infrared light of frequency $\omega = 920$cm$^{-1}$ onto the tip, detected the backscattered light, and isolated its near-field component $s_{nf}$ (*30–32*). A representative $s_{nf}$ image (Fig. 1B) demonstrates contrasting domains indicating differences in their optical conductivities. Since the optical conductivity $\sigma_{TBG}(\theta, \omega)$ of TBG is sensitive to $\theta$ (*33*, *34*), we interpret these domains as regions of distinct twist angles, similar to the domains previously observed by STM (*10*).

Next, we performed photocurrent nanoscopy. Figure 1C depicts $I_{PC}$ measured in the same region as Fig. 1B. We observe two key features. First, the magnitude of the photocurrent signal is enhanced when the tip is located over domain boundaries. The magnitude of $I_{PC}$ varies along the boundaries, as most clearly seen for the boundary between monolayer graphene (MLG) and Region B. Second, we observe varying levels of photocurrent within Region A. In Fig. 1D we plot the photocurrent along the dashed line in Fig 1C. The increase in the magnitude of photocurrent when the tip is above the boundaries leads to minima with widths of about 200 nm each (half width at half minimum) while the remaining features correspond to local variations in the photocurrent signal within the domain.

In general, photocurrent arises from an interplay of optical and transport phenomena that occur at different temporal and spatial scales. Two important length scales in our case are the tip radius and the cooling length. The former, of the order of 10 nm, sets the size of the field enhancement region where non-equilibrium charge carriers are generated and is also the spatial resolution of infrared nanoscopy (*35*). The latter can be a few hundred nanometers or longer in graphene, depending on the experimental conditions (*27*, *36*) and determines the size of the "hot spot" around the tip where the electron temperature remains elevated. Assuming that photocurrent is predominantly due to the photo-thermoelectric effect (PTE) (*28*, *36*), the photocurrent scales approximately as $I_{PC} \propto \Delta T \Delta S$, where $\Delta T$ is the change in electron temperature induced around the tip and $\Delta S$ is the change in the local Seebeck coefficient $S$ across the hot spot, whose direction determines that of the current flow. The photocurrent generated from the PTE also varies on the length scale of the cooling length. Therefore, photocurrent data represent a coarse-grained measurement of gradients in Seebeck coefficient. We observe this effect as enhanced $I_{PC}$ at MLG-TBG and TBG-TBG boundaries due to a discontinuity $\Delta S$ of the Seebeck coefficient across such boundaries. The slower variation in $I_{PC}$ along the boundaries is due to the changing direction of the current flow which is dictated by the geometry of the electrical contacts (Section S3 of (*29*)). The remaining short-range variations of $I_{PC}$ seen in Fig 1D are attributed to short-range variations of the optical conductivity. These latter contrasts likely arise from a combination of charge puddles and twist angle variations (*20*). Further experiments are needed to distinguish between these possibilities.

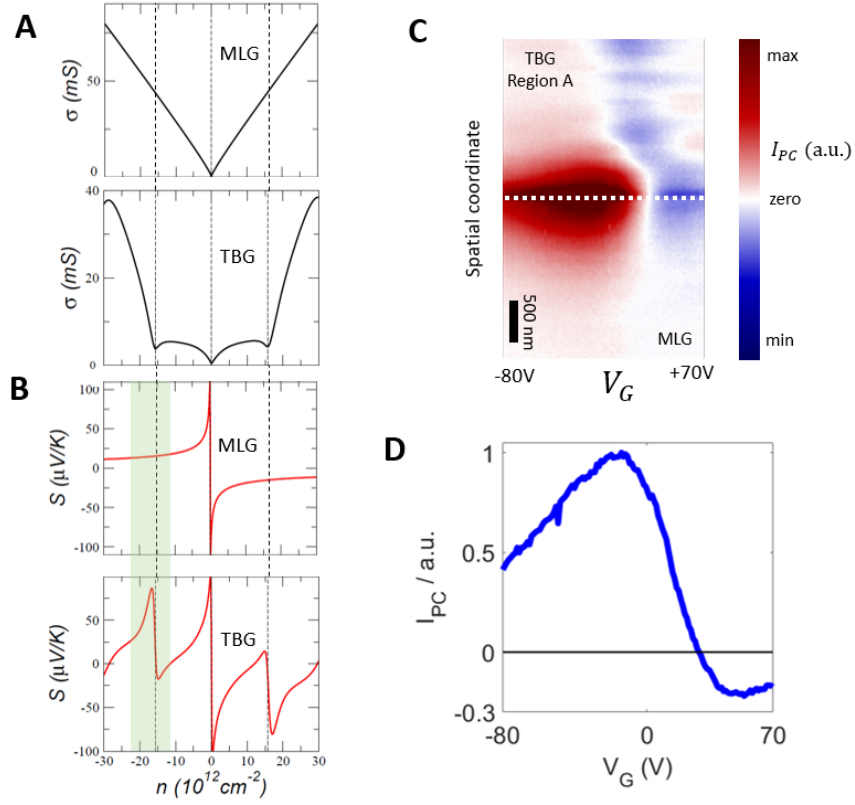

**Figure 2 | Photocurrent spectroscopy at the interface between twisted bilayer graphene and monolayer graphene.** (A) Calculated DC conductivity of MLG and TBG with $\theta = 2.65°$. (B) Calculated Seebeck coefficients for MLG and TBG with $\theta = 2.65°$. The dashed lines correspond to the superlattice band edges and the green area corresponds to the experimentally measured range of densities in (C) and (D). (C) A nano-photocurrent carrier density sweep across a MLG-Region A interface acquired at $T = 200K$. (D) Gate voltage dependence of the photocurrent at the MLG-Region A boundary at $T = 200K$ (white dashed line in (C)).

We now elucidate the ability of the nano-photocurrent method for characterizing the TBG domains using gate voltage dependent measurements. As pointed out above, the photocurrent near a TBG-MLG boundary is proportional to the difference in the Seebeck coefficient across the boundary: $I_{PC} \propto \Delta S = S_{TBG} - S_{MLG}$, where $S_{TBG}$ and $S_{MLG}$ are the Seebeck coefficients for TBG and MLG respectively. In this work, we will neglect the correlated electron physics in the moiré flat bands (*6, 19*): a valid assumption for all twist angles away from the magic angle. Within this assumption, the Seebeck coefficient obeys the Mott formula for both MLG and TBG (Section S6 of (*29*), (*37*)):

$$S = -\frac{\pi^2 k_B^2 T_{el}}{3|e|} \frac{1}{\sigma} \frac{d\sigma}{d\mu} \qquad (1)$$

where $\sigma$ is the electrical conductivity of graphene, $\mu$ is the chemical potential, and $T_{el}$ is the electronic temperature. The DC conductivity of MLG, $\sigma_{MLG}$ shows a symmetric dip at the charge neutrality point (CNP) $n = 0$ (Fig. 2A, top). In turn, $S_{MLG}$ is an odd function of the carrier density with peaks above and below the CNP (*38*) (Fig 2B, top). Similarly, the calculated DC conductivity of TBG $\sigma_{TBG}$ (Fig. 2A bottom) indicates that $\sigma_{TBG}$ has three minima: one at the CNP and two more at carrier densities $n = \pm n_s$ associated with superlattice band edges (*5*). The Mott formula then predicts that the Seebeck coefficient $S_{TBG}$ of TBG should exhibit characteristic zigzag-like variations versus carrier density close to $n = \pm n_s$ (Fig. 2B, bottom). We note that direct measurement of the Seebeck coefficient of TBG through conventional thermoelectric measurements should also reveal the characteristic zigzag pattern.

For moderately small twist angles, the analysis of the photocurrent signal can be further simplified. For $\theta > 1°$, $|n_s| > 3 \cdot 10^{12} \text{cm}^{-2}$, the Seebeck coefficient of MLG at $n = \pm n_s$ can be neglected in a first approximation compared to that of TBG (*38*, *39*), so that $I_{PC} \propto S_{TBG}$. Since $S_{TBG}$ changes sign at $n = \pm n_s$, the photocurrent $I_{PC}$ is also expected to change sign as well. Detection of such a zero crossing of $I_{PC}$ measured locally as a function of $n$ can then be used to estimate $n_s$ and thereby the twist angle of TBG. We denote this estimate by $\theta_{PC}$.

Nano-photocurrent experiments are robust to device architecture and disorder effects. The existence of the zero crossing in $I_{PC}$ does not depend on geometric factors such as the position of the electrical contacts used to measure $I_{PC}$ or the size and relative location of the TBG domains. Therefore, we can study multiple twist angle domains across the device with a single pair of electrical contacts. Further modelling developed in Section S4 of (*29*) shows that the sign change is also insensitive to fine details such as disorder strength which affects $\sigma_{MLG}$ and $\sigma_{TBG}$ and a finite value of $S_{MLG}$ at $n = \pm n_s$, so long as $\sigma_{TBG}$ exhibits a minimum at $n = \pm n_s$.

Since the SBEs in TBG are observable only at cryogenic temperatures (*18*), we performed nano-photocurrent experiments in a home-built ultra-high-vacuum platform for low temperature nano-imaging (*40*). Figures 2C and 2D show the results of such an experiment across a MLG/TBG interface at $T = 200$K. We see that the interface serves as a strong source of photothermoelectric current when the near-field probe is brought within proximity of a few hundred nanometers. Recording $I_{PC}$ while scanning repeatedly across this interface and changing the bias $V_g$ applied to the gate electrode produces a spatial map of photocurrent at different carrier densities. For this particular MLG/TBG interface, we find a sign change in $I_{PC}(n)$ at a density $n_A = -1.58 \cdot 10^{13} \text{cm}^{-2}$ corresponding to $\theta_{PC} = 2.61°$ (Section S2 in (*29*)). Further evidence for the presence of a SBE at this specific carrier density comes from the observation of line-like features in $I_{PC}$ close to the sign change that only appear in the TBG region. Such features, previously observed in MLG and BLG close to CNP (*41*, *42*), can arise from spatial inhomogeneities in carrier density (*23*, *41*) as well as local variations in the twist angle (*20*) which lead to comparatively large spatial variations in Seebeck coefficient and serve as local sources of photocurrent. In our data, these features are spatially confined to the TBG region and only appear when $n$ is close to $n_A$. The totality of these observations suggest that the Seebeck coefficient in

TBG reveals spatial variations most prominently for carrier densities $n \approx n_A$, thereby confirming the presence of the SBE in Region A at $n = n_A$.

We applied the same protocol of nano-photocurrent imaging and gate sweeps at the Region B-MLG interface (Fig 1C). For Region B, we found no sign change at either $T = 200K$ or $T = 40K$ (data shown in Section S5 of (29)). By constraining the densities at which the first order and second order superlattice band edges appear (Section S5.2 in (29)), we conclude that $2.27° < \theta_{PC} < 2.34°$ for Region B. We have, therefore, measured the twist angle for two different regions of our device through nano-photocurrent with the same pair of electrical contacts.

The large carrier density and limitations of the back-gate in our device prevented us from reaching the CNP or the SBEs for electron-doped Fermi levels in TBG. To confirm that our assignment of $\theta_{PC}$ is accurate, we performed nano-infrared imaging experiments. In nano-IR experiments, infrared light incident on the metallic tip launches surface plasmon polaritons in graphene (43, 44) which are reflected by physical (45, 46) or electronic boundaries (47–49) and form standing wave patterns that can be directly imaged (46, 50). The wavelength and the spatial decay length of the plasmons are directly related to the optical conductivity of the material at the energy of the incident light (45, 46, 51).

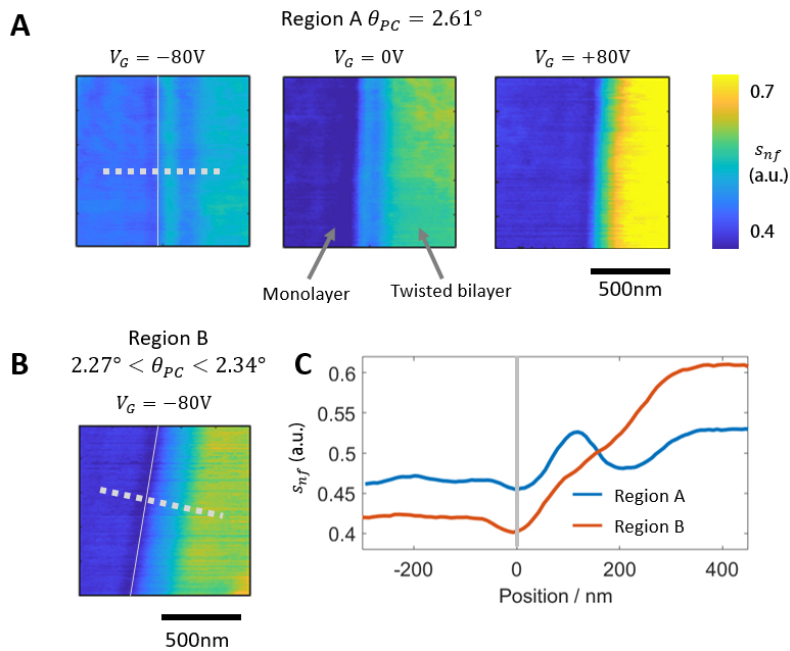

**Figure 3 | Nano-infrared images of MLG-TBG interface.** (A) Amplitude of the backscattered light $s_{nf}$ images of Region A at three different values of $V_G$, demonstrating the gate voltage dependent plasmonic properties. (B) $s_{nf}$ image of Region B at $V_G = -80V$. (C) Line profiles along the grey dashed lines in Fig 2A and 2B illustrating the difference in the plasmonic fringes. Grey solid lines indicate the MLG-TBG boundary. All images were acquired at $T = 300K$ with infrared light of frequency $\omega = 920 cm^{-1}$.

Our nano-infrared imaging data taken at a MLG-TBG interface for Regions A and B are shown in Fig 3. We observe strong fringes on the TBG side of the MLG-TBG boundary, as seen in the line profiles in Fig 3C, which we assign to plasmons propagating in TBG that are reflected by the MLG-TBG boundary. Plasmons propagating in MLG are also visible as faint fringes on the MLG side. The fringes in TBG are gate tunable (Fig 3A) and are strongly dependent on the twist angle (Fig 3B, 3C), resulting from the optical conductivity of TBG being highly sensitive to the carrier density and twist angle.

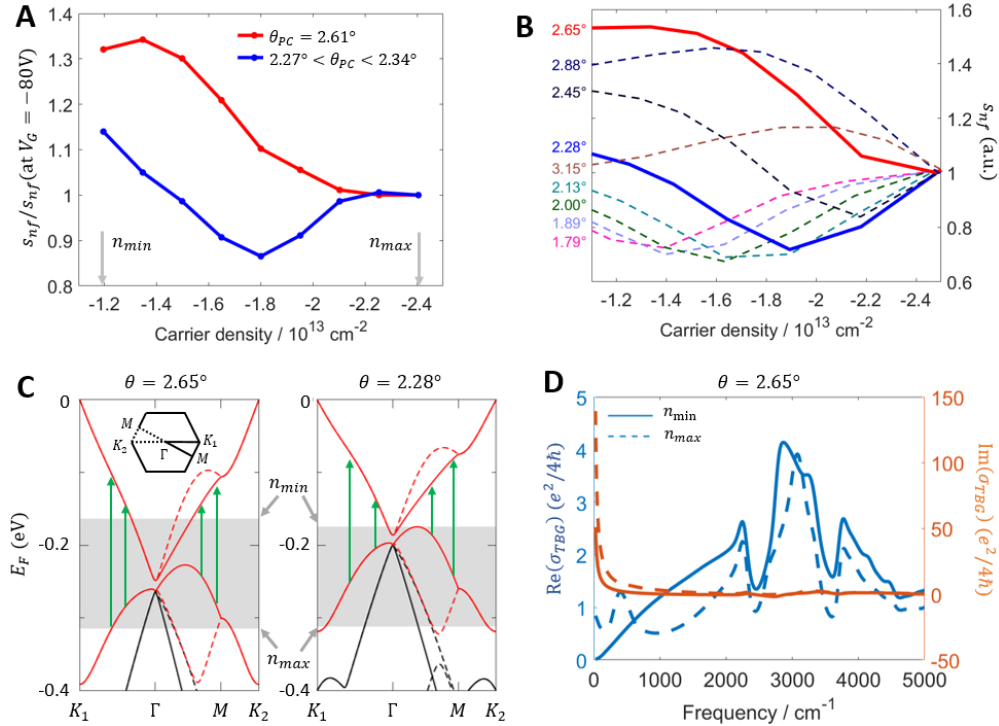

**Figure 4 | Optical response of TBG at small angles.** (A, B) $s_{nf}$ as a function of carrier density in experiment (A) and in calculations based on the continuum model (B) at $\omega = 920 \text{cm}^{-1}$. (C) Band structures for two different twist angles illustrating the twist angle dependent changes. Grey areas represent the range of carrier densities accessed in the experiment. The green arrows represent the optical transitions that are suppressed as the carrier density increases. Solid and dashed lines correspond to the solid and dashed paths through the moiré Brillouin zone as shown in the inset. (D) Real and imaginary parts of the optical conductivity calculated for $\theta = 2.65°$ at $n_{min} = -1.10 \cdot 10^{13} \text{cm}^{-2}$ (solid) and $n_{max} = -2.48 \cdot 10^{13} \text{cm}^{-2}$ (dashed).

We now compare our nano-infrared imaging data to theoretical models. We average the measured amplitude of the back scattered light $s_{nf}$ in the TBG regions and plot it as a function of carrier density. We calculated the optical conductivity of TBG $\sigma_{TBG}(\theta, n)$ at $\omega = 920 \text{cm}^{-1}$ using the continuum model (*1*, *29*) for the nine commensurate angles between 1.79° and 3.15°. We then applied a theoretical model of tip-sample coupling (*52*) to calculate $s_{nf}(\theta, n)$ from $\sigma_{TBG}(\theta, n)$ (*29*). The results of our calculations are shown in Figure 4B. The experimentally measured near-field optical response recorded from TBG in Regions A and B (Fig. 4A) are in excellent agreement with the calculations, thereby validating our estimates of the twist angle deduced from photocurrent nanoscopy.

The features observed in $s_{nf}(n)$ provide further insight into the band structure of TBG. The decrease in $s_{nf}(n)$ for $\theta = 2.65°$ and the dip for $\theta = 2.28°$ for increasing $n$ indicate the presence of optical transitions that are being tuned by carrier density. Comparison with the calculated band structures for TBG at the associated twist angles demonstrates this is indeed the case. Figure 3(C) presents the band structure for TBG for $\theta = 2.65°$ and $\theta = 2.28°$. The Fermi energy $E_F$ range accessible in our experiment is shown in gray, where the upper boundary corresponds to $n_{min} = -1.1 \cdot 10^{13}$ cm$^{-2}$ and the lower boundary corresponds to $n_{max} = -2.48 \cdot 10^{13}$ cm$^{-2}$. We see that towards the higher range of attainable $E_F$, transitions between the minibands such as those shown by the green arrows are suppressed. This suppression leads to a decrease in the real part of the optical conductivity at the probing frequency of 920cm$^{-1}$ (Figure 3D) and a corresponding drop in $s_{nf}(n)$ for $\theta = 2.65°$. For $\theta = 2.28°$, $s_{nf}(n)$ begins to increase at $n < -1.8 \cdot 10^{13}$cm$^{-2}$ as transitions between other minibands begin to contribute (Section S6 in (*29*)).

The cryogenic photocurrent nanoscopy technique utilized here can be applied broadly to characterize the electronic structure of TBG and its variation across macroscopic structures. Photocurrent nanoscopy can also be applied to moiré patterns in other vdW heterostructures. When graphene is placed on hBN, twist-angle dependent SBEs, similar to those in TBG, appear (*53*, *54*). Photocurrent nanoscopy is well suited to resolve local changes in the twist angle in such structures. Domains of different stacking orders in multilayer graphene such as ABC and ABA stackings in trilayer graphene (*55*) can also be probed with nano-photocurrent imaging. Recently, evidence for local variations of the excitonic properties due to moiré patterns in transition metal dichalcogenide (TMD) bilayers (*56*–*59*) has been reported. While far-field photocurrent experiments have demonstrated sensitivity to the exciton resonance in TMDs (*60*), cryogenic photocurrent nanoscopy needs to be applied to resolve sub-micron changes in the excitonic properties of TMD bilayers.

To conclude, we have demonstrated the utility of nano-photocurrent technique to locally probe the electronic structure of small-angle TBG and to determine its twist angle with nano- and mesoscale spatial resolution. The technique detects local variations in the photothermoelectric effect in graphene, which is highly sensitive to formation of a superlattice band edge in TBG, thus providing a fast, robust and transport-compatible method for evaluating the twist angle of TBG. Photocurrent nanoscopy does not require any special device architectures and only necessitates optical access to the graphene layers together with a pair of global electrical contacts.

Photocurrent nanoscopy can also be extended to characterize the electronic structure of other van der Waals heterostructures such as multilayer graphene and TMDs.

Acknowledgements: Research on charge dynamics of graphene at Columbia is supported by DOE-BES DE-SC0018426. Research at UCSD is supported by ONR-N000014-18-1-2722. Development of novel nano-imaging instrumentation at Columbia is supported ONR-N000014-18-1-2722 and AFOSR: FA9550-15-1-0478. Work on van der Waals photonic structures is supported AFOSR: FA9550-15-1-0478. TS acknowledges support from Spain's MINECO under Grant No. FIS2017-82260-P as well as by the CSIC Research Platform on Quantum Technologies PTI-001. PK and HY acknowledge support from ONR MURI on Quantum Optomechanics (Grant No. N00014-15-1-2761).

The Supporting Information contains Materials and Methods, Carrier density and superlattice band edge calculations, Large area nano-infrared and nano-photocurrent images, Simple model for photocurrent at a MLG-TBG interface, Photocurrent data for other twist angles and temperatures, Conductivity and Seebeck coefficient calculations and Lightning rod model calculations

# Supporting Information for "Nano-photocurrent mapping of local electronic structure in twisted bilayer graphene"

Sai S. Sunku[1,2]*, Alexander S. McLeod[1],*, Tobias Stauber[3], Hyobin Yoo[4], Dorri Halbertal[1], Guangxin Ni[1], Aaron Sternbach[1], Bor-Yuan Jiang[5], Takashi Taniguchi[6], Kenji Watanabe[6], Philip Kim[4], Michael M. Fogler[5], Dimitri N. Basov[1],†

1. Department of Physics, Columbia University, New York, NY

2. Department of Applied Physics and Applied Mathematics, Columbia University, New York, NY

3. Departamento de Teoría y Simulación de Materiales, Instituto de Ciencia de Materiales de Madrid, CSIC, Madrid, Spain

4. Department of Physics, Harvard University, Cambridge, MA

5. Department of Physics, UC San Diego, La Jolla, CA

6. National Institute for Materials Science, Tsukuba, Japan

* - These authors contributed equally

† - Corresponding author db3056@columbia.edu

## S1: Materials and Methods

<u>Device fabrication</u>

Our device consists of twisted bilayer graphene fabricated using the tear-and-stack technique with the graphene directly on $SiO_2$. First a layer of boron nitride (BN) is picked up using an adhesive polymer poly(bisphenol A carbonate) (PC) coated on a stamp made of transparent elastomer polydimethylsiloxane (PDMS). A large flake of monolayer graphene is identified and the BN flake is used to tear the graphene flake into two and pick up one half. The substrate is then rotated by a controlled angle and the second half of the graphene flake is picked up. The entire stack is then placed directly on a silicon dioxide/silicon substrate without a bottom BN layer. The presence of dopants on the $SiO_2$ surface leads to a high carrier density in graphene even without the application of a gate voltage. The same device was investigated in our previous work on photonic crystals (S*1*).

<u>Infrared nano-imaging</u>

Infrared nano-imaging was performed with a commercial scattering-type scanning near-field optical microscope (s-SNOM) based on a tapping mode atomic force microscope from Neaspec GmbH. Our light source was a quantum cascade laser obtained from DRS Daylight Solutions, tunable from 900 cm$^{-1}$ to 1200 cm$^{-1}$. The light from the laser was focused onto a metallic tip oscillating at a tapping frequency of around 250 kHz with a tapping amplitude of around 60 nm. The scattered light was detected using a liquid nitrogen cooled HgCdTe (MCT) detector. To suppress far-field background signals, the detected signal was demodulated at a harmonic $n$ of the tapping frequency. In this work, we used $n = 3$.

Nano-photocurrent

Room temperature nano-photocurrent measurements were performed in a commercial s-SNOM from Neaspec GmbH. Low temperature nano-photocurrent measurements were performed in a home-built ultrahigh vacuum chamber (S2). The incident laser power was around 20mW for room temperature experiments and 40mW for low temperature experiments. The current was measured using a Femto DHPCA-100 current amplifier. To isolate the photocurrent contributions from the near-fields localized under the tip, the measured current was demodulated at a harmonic $n$ of the tapping frequency. In this work we used $n = 2$. The gate sweeps were performed by scanning the same line repeatedly while slowly changing the gate voltage. We typically swept the gate voltage at a rate of 100mV/sec while the time required to scan a single line was about 10 seconds.

## S2: Carrier density and SBEs
### S2.1: Estimating the carrier density in our device

We estimate the carrier density in two ways: plasmon wavelength in monolayer graphene and Raman experiments. Both experiments confirm that the carrier density at $V_G = 0$V is $\sim -1.8 \cdot 10^{13}$ cm$^{-2}$. This extremely large carrier density is likely the result of doping from the SiO$_2$ surface that the graphene directly sits on. In the room temperature nano-infrared experiments, we were able to apply gate voltages between $V_G = -80$V and $V_G = +80$V. However for the low temperature nano-photocurrent experiments, we could only reach $V_G = +70$V.

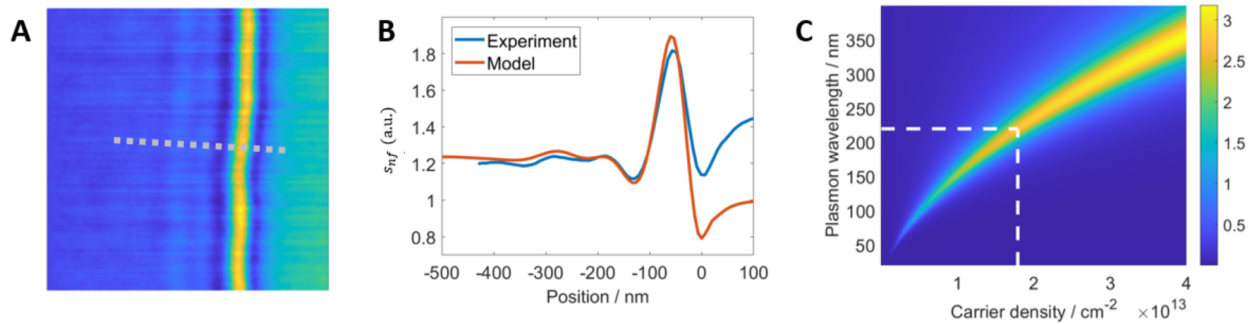

**Figure S1 | Nano-infrared image of MLG used to estimate the carrier concentration.** (A) Nano-infrared image acquired at a MLG edge. (B) Line profile perpendicular to the edge along with a fit using $\lambda_p = 220$nm. (C) Plasmon dispersion for the BN/MLG/SiO2/Si heterostructure. A plasmon wavelength of 220nm corresponds to a carrier concentration of $1.78 \cdot 10^{13}$cm$^{-2}$.

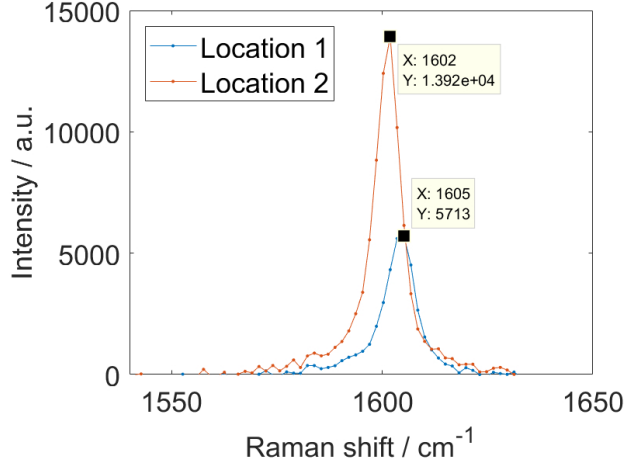

**Figure S2 | Raman measurements on MLG at $V_G = 0V$.** The position of the G peak is around 1603cm$^{-1}$ which corresponds to a carrier density of $\sim -1.75 \cdot 10^{13}$cm$^{-2}$ (*S3*).

### S2.2: Carrier density of SBEs

The carrier density at which SBEs occur can be calculated as follows. The first SBE appears when the first moiré band is filled. Since the band structure of TBG is four-fold degenerate (including spin and valley), this corresponds to 4 carriers per moiré unit cell. The area of the moiré unit cell is given by:

$$A_M = \frac{\sqrt{3}}{2} \lambda_M^2 \tag{S1}$$

Therefore, the density at the SBE, $n_s$ is given by:

$$n_s = \frac{4}{A_M} = \frac{4}{\frac{\sqrt{3}}{2} \lambda_M^2} = \frac{4}{\frac{\sqrt{3}}{2} \left(\frac{a}{2}/\sin\left(\frac{\theta}{2}\right)\right)^2} \tag{S2}$$

where $a$ = 0.246nm is the lattice constant of monolayer graphene. Eq S2 provides a direct relationship between $n_s$ and $\theta$ and allows for a determination of $\theta$ from a measurement of $n_s$.

## S3: Large area nano-infrared and nano-photocurrent images of our sample

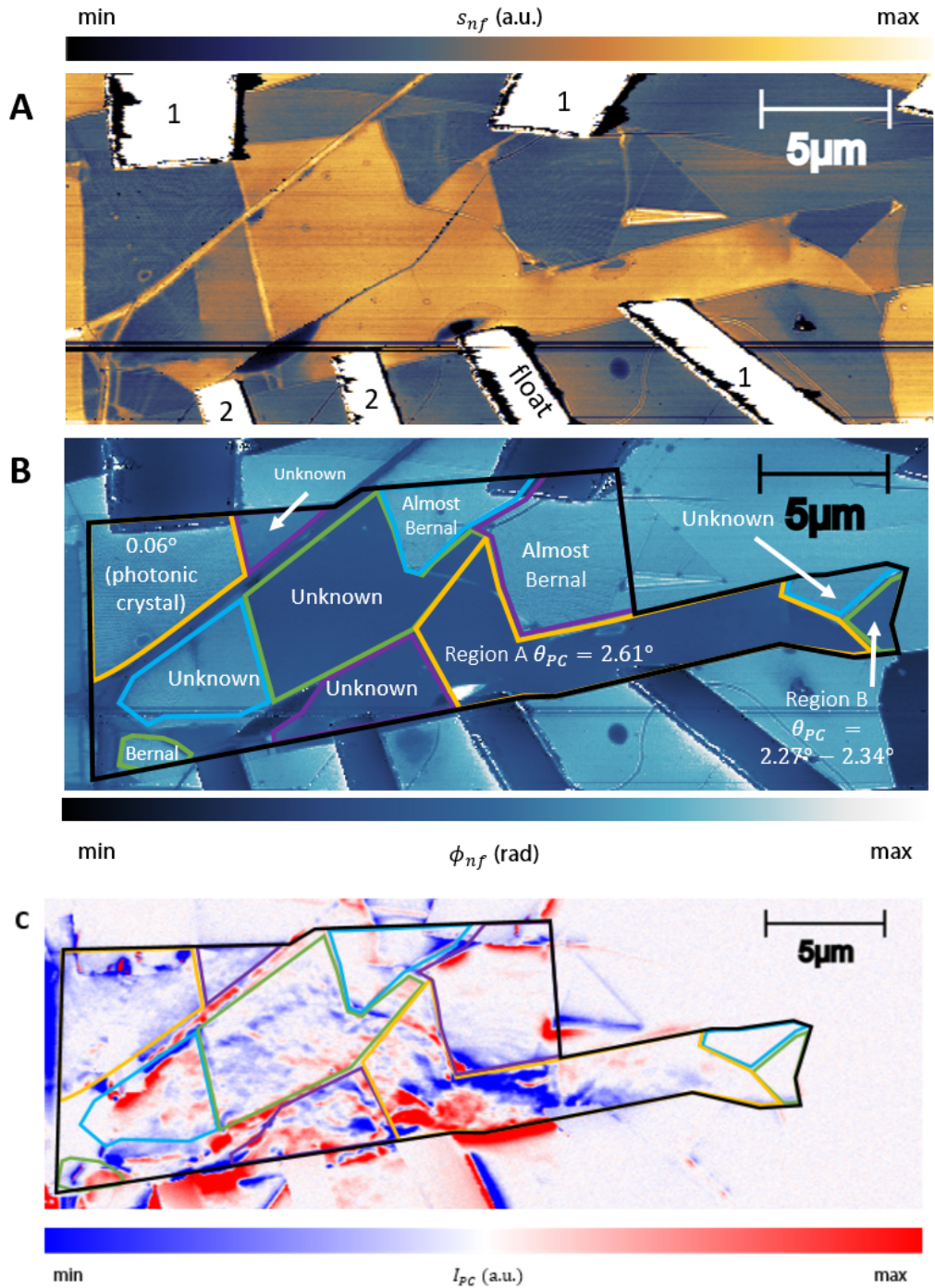

**Figure S3 | Large area images of our device.** (A) Nano-infrared image of the amplitude $s_{nf}$. The bright white regions are the gold electrodes and the numbers indicate the electrode configuration used for photocurrent experiments. (B) Nano-infrared image of the phase $\phi_{nf}$. The various regions of the sample are marked. 'Unknown' refers to regions that we were unable to determine the twist angle conclusively. (C) Nano-photocurrent image. The colored lines in (B) and

(C) correspond to the boundaries of domains with different twist angles. All images were acquired at $V_G = 0V$ and at room temperature.

Large area nano-infrared and nano-photocurrent images are shown in Figure S3. The electrode configuration for nano-photocurrent experiment in shown in Fig S3(A). The electrodes labeled '1' and '2' served as source and drain respectively. The electrode labeled 'float' was left floating. Previous work on photocurrent generation in graphene has shown that the measured geometric pattern of the photocurrent is strongly sensitive to the geometry of the electrical contacts (S4). The unusual geometry used here results is responsible for the variations of the measured signal across the boundaries, as seen in Figure 1(C).

### S4: Simple model for photocurrent at a MLG-TBG interface

In the following, we develop a simple model for the photocurrent generated at a MLG-TBG interface. First, we assume that the resistivity of MLG can be parametrized by a single quantity $n_{dis}$ which is the full-width half-maximum (FWHM) of the resistivity as a function of carrier density and is a measure of the disorder in the device:

$$R_{MLG} = \left(1 + \frac{n^2}{n_{dis}^2}\right)^{-1}$$

For TBG, we can consider three contributions: one from the CNP and two from the SBEs:

$$R_{TBG} = \left(1 + \frac{n^2}{n_{dis}^2}\right)^{-1} + \left(1 + \frac{(n - n_s)^2}{n_{dis}^2}\right)^{-1} + \left(1 + \frac{(n + n_s)^2}{n_{dis}^2}\right)^{-1}.$$

We can then use the Mott formula to calculate the Seebeck coefficient for MLG and TBG and use that to estimate the difference between the actual twist angle $\theta$ and the twist angle estimated from the nano-photocurrent measurements $\theta_{PC}$. Figure S4(A) shows an example of the Seebeck coefficient curves that result from the above model. Figure S4(B) shows the relationship between $|\theta - \theta_{PC}|$ as a function of $\theta$ and $n_{dis}$. From Fig S4(B), we see that the difference between $\theta$ and $\theta_{PC}$ is significant only when $n_{dis}$ exceeds $\sim 10^{11} cm^{-2}$. The latest generation of high quality encapsulated graphene devices show $n_{dis} \sim 10^9 cm^{-2}$ (S5). Furthermore, at small twist angles, the SBEs show finite electronic gaps (S6, S7) which would lead to sharper changes in the resistance and Seebeck coefficient as a function of the density close to the SBEs, leading to a smaller difference between $\theta$ and $\theta_{PC}$. Therefore, for practical experiments, $\theta$ and $\theta_{PC}$ can be considered to be identical.

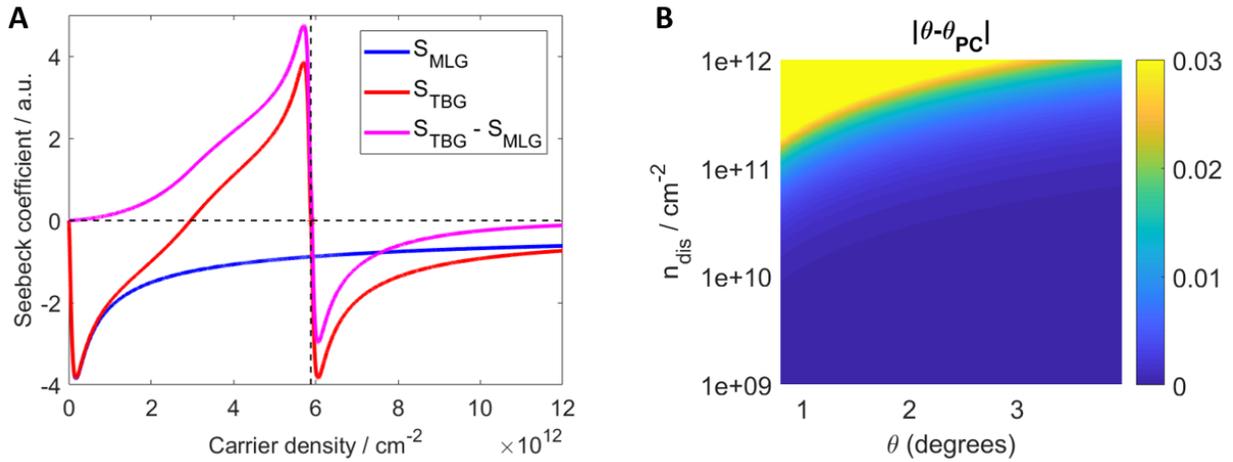

**Figure S4 | Simple model for photocurrent at a MLG-TBG interface.** (A) Seebeck coefficient for MLG, TBG and their difference as a function of carrier density. The parameters used are $\theta = 1.59°$ and $n_{dis} = 1 \cdot 10^{11} \text{cm}^{-2}$. The horizontal dashed line represents $S = 0$ and the vertical dashed line represents the SBE for the TBG. (B) A color plot of $|\theta - \theta_{PC}|$ as a function of $\theta$ and $n_{dis}$. A significant difference between $\theta$ and $\theta_{PC}$ is only observed for $n_{dis} > 10^{11} \text{cm}^{-2}$.

## S5: Photocurrent data for other twist angles and temperatures

S5.1 Nano-photocurrent data for Region A $\theta_{PC} = 2.61°$ at $T = 40K$

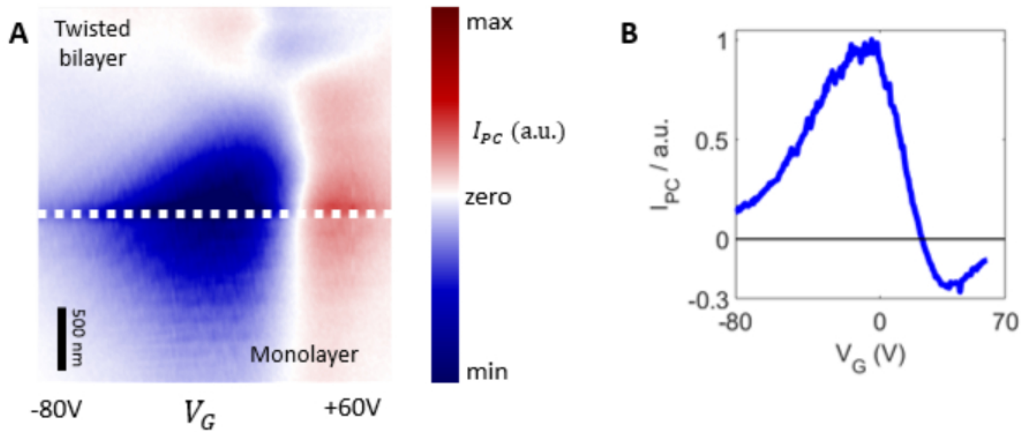

**Figure S5 | Nano-photocurrent data for the $\theta_{PC} = 2.61°$ region at $T = 40K$.** (A) Nano-photocurrent image. (B) Line profile along the white dashed line in (A). The sign change in $I_{PC}$ occurs at a density of $-1.62 \cdot 10^{13} \text{cm}^{-2}$ which is very similar to the sign change density at $T = 200K$ of $-1.58 \cdot 10^{13} \text{cm}^{-2}$ (shown in Figure 1 of the main text).

## S5.2 Nano-photocurrent data for Region B $2.27° < \theta_{PC} < 2.34°$

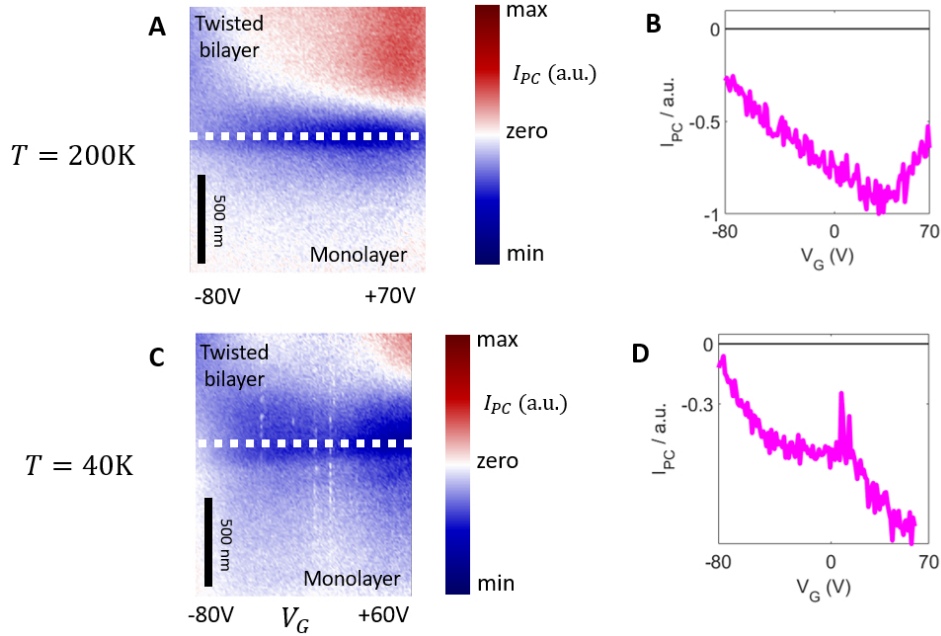

**Figure S6 | Nano-photocurrent data for the $2.27° < \theta_{PC} < 2.34°$ region** (A) Nano-photocurrent image and (B) Line profile at $T = 200$K. (C) Nano-photocurrent image and (D) Line profile at $T = 40$K.

For this region, we do not observe a sign change in $I_{PC}$. However, we can place bounds on the twist angle by considering the range of carrier densities accessible in our experiment. There are two possibilities. In the first possibility, the twist angle is so large that the SBE density is greater than $n_{max}$. This requirement leads to a bound of $\theta_{PC} > 3.21°$.

The second possibility is that the twist angle is so small that the SBE density is below $n_{min}$. This requirement leads to a bound of $\theta_{PC} < 2.34°$. However, a lower bound may also be established in this case by requiring that the second order SBE (observed at smaller angles in (S8, S9) and predicted by our calculations in Figure S8, Section S6) be above $n_{max}$, leading to the bound $\theta_{PC} > 2.27°$. Combining the two bounds, we get $2.27° < \theta_{PC} < 2.34°$. Because this second possibility is a more stringent requirement on $\theta_{PC}$, we use this requirement in the main text. The good agreement between the nano-infrared signal in Fig 3 is further confirmation that our estimate of $\theta_{PC}$ is accurate.

In the $T = 40$K data for Region B, we observe a suppression in photocurrent at $V_G \sim 12$V. We believe that this effect is likely the result of Region A being at the SBE. Since Region B is located away from the electrodes, the photocurrent must flow through Region A to reach the electrodes. If Region A were to become gapped, it will lead to an apparent suppression of the photocurrent from Region B.

## S5.3 Real space nano-photocurrent images of Region A

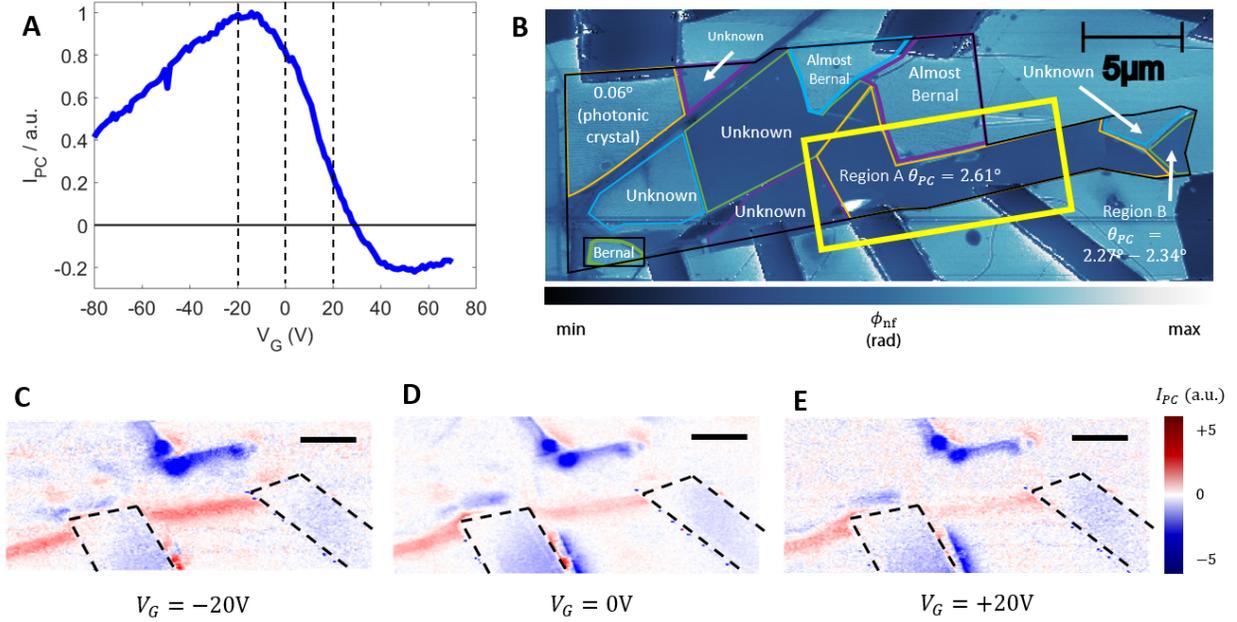

**Figure S7 | Real space nano-photocurrent images of Region A and its boundary with MLG.** (A) Gate voltage dependence of the photocurrent at a Region A-MLG interface (same as Fig 1(E)). (B) Large area near-field phase image (same as Fig S3(B)). The yellow rectangle marks the region shown in panels C-E. (C-E) Nano-photocurrent image at three different values of $V_G$. The photocurrent at the boundary decreases as $V_G$ changes from -20V to +20V as expected from the voltage dependence in panel A. Panel A was acquired at $T = 200$K and panels C-E were acquired at $T = 160$K.

## S6: Conductivity and Seebeck coefficient calculations

The DC conductivity and Seebeck coefficient of MLG and TBG were calculated using semiclassical Boltzmann theory (*S10*, *S11*) within the continuum model. We assumed an energy dependent scattering rate originating from Coulomb interactions as well as short-ranged scatterers and used Eq 2 and Eq 3 from Ref (*S10*). For the temperature range used in our experiments, the Seebeck coefficient can be obtained from the conductivity as function of the chemical potential using the Mott formula (Eq 9 of (*S10*)) thus justifying the use of Mott formula also in TBG (Figure S8).

Optical conductivity of TBG was calculated again using a continuum model. First, the real part of the optical conductivity was obtained by replacing the delta-function by a Gaussian with variance of 3 meV. The imaginary part was then obtained from the Kramers-Kronig relation after adding the constant background $\sigma_0 = e^2/2\hbar$ for frequencies larger the cut-off frequency set by the continuum model (S*12*). The Drude peak in the absorption was neglected due to the large relaxation times estimated to significantly exceed 1fs.

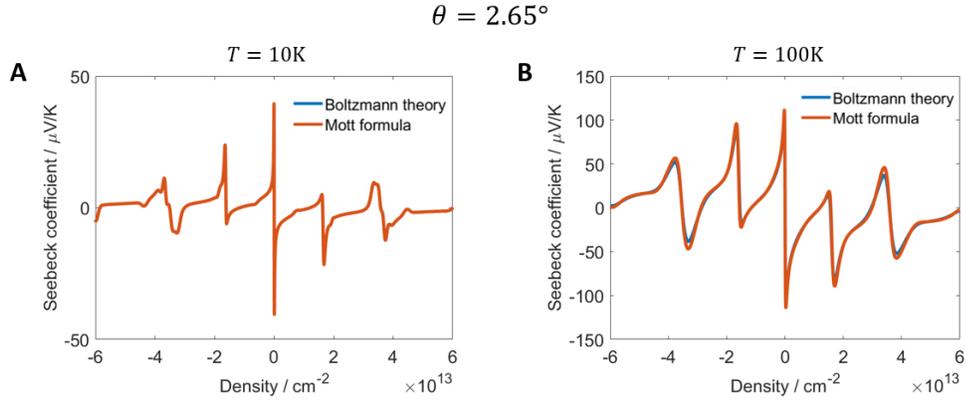

**Figure S8 | Comparison between Boltzmann theory and the Mott formula.** Seebeck coefficient of TBG with $\theta = 2.65°$ calculated using Boltzmann theory and the Mott formula at $T = 10K$ and $T = 100K$. The good agreement justifies the use of Mott formula in TBG.

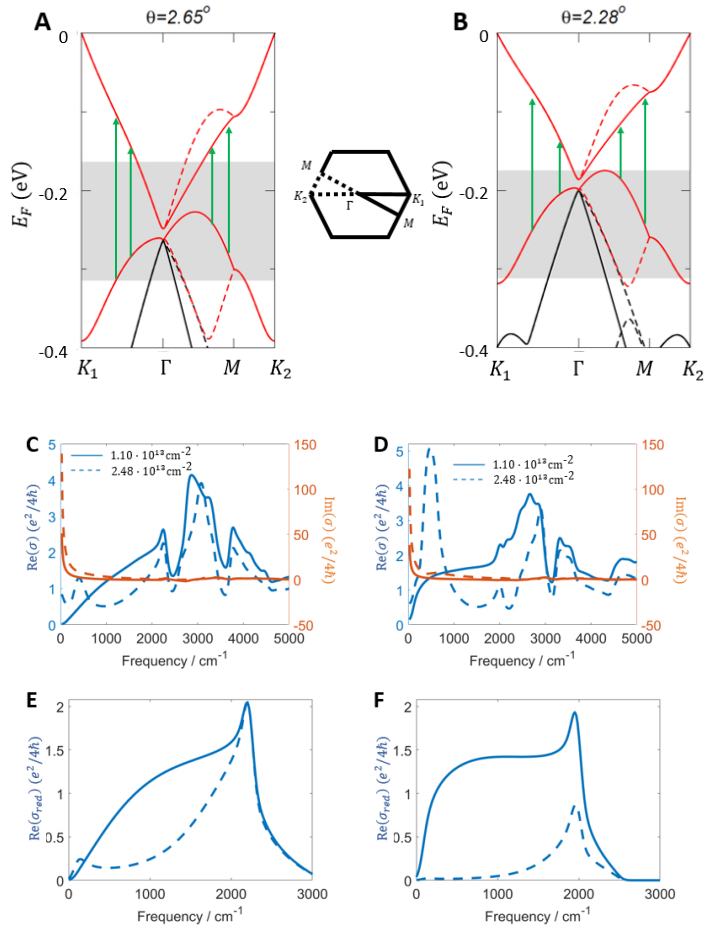

**Figure S9 | Optical conductivity spectra for different twist angles.** (A, B) Band structures for twisted bilayer graphene with $\theta = 2.65°$ and $\theta = 2.28°$ (same as Fig 4C in the main text). (C, D) Optical conductivity spectra for $\theta = 2.65°$ and $\theta = 2.28°$ at $n = 1.1 \cdot 10^{13} \text{cm}^{-2}$ (solid) and $n = 2.48 \cdot 10^{13} \text{cm}^{-2}$ (dashed). (E, F) $\text{Re}(\sigma_{red})$ spectra including only the transitions between the red bands in (C, D).

Figure S9 shows the optical conductivity spectra of TBG with $\theta = 2.65°$ and $\theta = 2.28°$. Fig S9(E) and S9(F) show the real part of the optical conductivity $\text{Re}(\sigma_{red})$ obtained by considering only the two red bands in Fig S9(A) and S9(B). $\text{Re}(\sigma_{red})$ is suppressed at $n = 2.48 \cdot 10^{13} \text{cm}^{-2}$ as compared to $n = 1.10 \cdot 10^{13} \text{cm}^{-2}$ because the red bands are no longer occupied. However, $\text{Re}(\sigma)$ for $\theta = 2.28°$ still shows a strong peak at 500cm$^{-1}$ at the higher density. This resonance does not arise from the red bands, but is the result of transitions from other moiré-modified bands at more negative Fermi energies.

### S7: Lightning rod model

We used the lightning rod model (LRM) (S13) to calculate the expected near-field amplitude of our hBN/TBG/SiO$_2$/Si heterostructure. The inputs for the LRM are the dielectric constants for each of the layers, the thicknesses of the layers and the radius of curvature of the tip. We used the output of the continuum model calculations for the conductivity of the TBG. For the dielectric constants of hBN and SiO$_2$, we used the parametrizations reported in Ref (S14) and Ref (S15) respectively. We estimated the radius of curvature of the tip to be around 15nm for our nano-infrared experiments.